\documentclass[aps,prl,reprint,groupedaddress]{revtex4-1}
\usepackage{graphicx,color}
\usepackage{amssymb}   % for math
\usepackage{amsmath,ulem,cancel}
\usepackage{epstopdf}
\usepackage[hidelinks]{hyperref}
\usepackage{bm}

\begin{document}
	
\title{Local Raman Spectroscopy of Chiral Majorana Edge Modes in Kitaev Spin Liquids and Topological Superconductors}

\author{James Jun He$ ^1 $, Naoto Nagaosa$ ^{1,2} $}
	
\affiliation{
		$ ^1 $ RIKEN Center for Emergent Matter Science (CEMS), Wako, Saitama 351-0198, Japan\\	
		$ ^2 $ Department of Applied Physics, The University of Tokyo, Tokyo 113-8656, Japan
	}
	
\date{\today}

\begin{abstract}

The Raman scattering with local optical excitation from the Majorana edge modes of Kitaev spin liquids and topological superconductors is studied theoretically. Although the effective one-dimensional model is common between these two cases, the coupling to the electromagnetic field is different. It is found that the Raman spectrum at low energy scales with $\omega^3$ in Kitaev spin liquids while it shows the gap in topological superconductors. This is in sharp contrast to the infrared absorption, where the spectrum shows the gap in Kitaev spin liquids, while it behaves as $\sim \omega^2$ in topological superconductors. This indicates that the electrodynamics of Majorana edge modes depends on their higher-dimensional origins. The realistic estimate of the Raman scattering intensity is given for $\alpha$-RuCl$_3$ as the candidate for Kitaev spin liquid.

\end{abstract}

\maketitle

Majorana fermions in condensed matter systems \cite{Read,Kitaev2001,Fu2008,Wilczek,Vic2009} attract  intensive interests recently. They are considered as a promising candidate for the robust quantum computation with the non-Abelian statistics \cite{Ivanov,Alicea2011,Fujimoto,SCZhangPNAS,Sato,Fujimoto,Nomura}. Robustness comes from the no or weak coupling with the surrounding systems.
Majorana fermions are neutral particles, and hence is intuitively expected to be decoupled from the electromagnetic field, which makes its observation difficult. 
For example, propagating Majorana modes are believed to exist in very different systems including quantum spin liquids (QSLs) \cite{Kitaev,Matsuda} and topological superconductors (TSCs) \cite{Fu2009,Beenakker2009,Balents,Qi2010,JJH_CP2019, Qinglin2017,ZWang,XGWen}. 
In QSLs, the Majorana fermions are neutral since they come from  spin operators while their neutrality in TSCs is manifested by their equal weight of electrons and holes at the zero energy. 
The origin of neutrality is that the Majorana field operator $ c(x) $ satisfies the relation $  c^\dagger(x) = e^{i \theta} c(x) $, with $ \theta $ being some phase, and hence the density operator $ c^\dagger(x) c(x) = e^{i\theta} c(x)^2$ is a numeric constant. 

However, 
it has been shown in a previous study \cite{JJH2021} that the optical response of Majorana edge modes in TSCs is nonzero. This is because of the existence of Cooper pair condensate. Then, it raises the question whether optical response also exists for the Majorana edge modes in QSLs or not. 
An analytically solvable model of QSLs is given by Kitaev \cite{Kitaev} which supports Majorana edge modes. There are two types of excitations in this model: vortices and Majorana fermions. The Raman scattering due to both of them  in the bulk has been investigated \cite{Knolle1,Nasu,Sandilands} while 
the infrared absorption is associated with both the vortices and Majorana fermions \cite{Bolens}. 
But the optical response of the edge states remains an open question. 
Particularly, optical microscopy and spectroscopy methods \cite{Jarzembski,Verma2017,Zhang2017} have been successful in detecting local optical responses with high spatial resolution (up to a few nanometers \cite{Lee,Klingsporn}), which is ideal for the investigation of topological edge states \cite{ZXShen,Lai}.

In this paper, we theoretically investigate the Raman scattering in a Kitaev QSL with a magnetic field and in a spinless $ p $-wave SC, both of which host gapless chiral Majorana edge channels while the bulk energy spectrum is gapped. 
In the QSL, we found that the edge channel induces Raman scattering with an intensity  proportional to $ \omega^3 $,  $ \omega $  being the photon frequency shift. We estimate the realistic value of the intensity and find it detectable with current local optical techniques.
In contrast, the edge channel in the $ p $-wave SC does not induce Raman scattering unless $ \hbar \omega $ is larger than the bulk gap. 
Such essential differences between our results for QSLs and TSCs indicate that there does not exist a general  effective edge theory which is sufficient to describe the electrodynamics of Majorana edge modes of various origins.

\begin{figure}
	\includegraphics[width=3.2in]{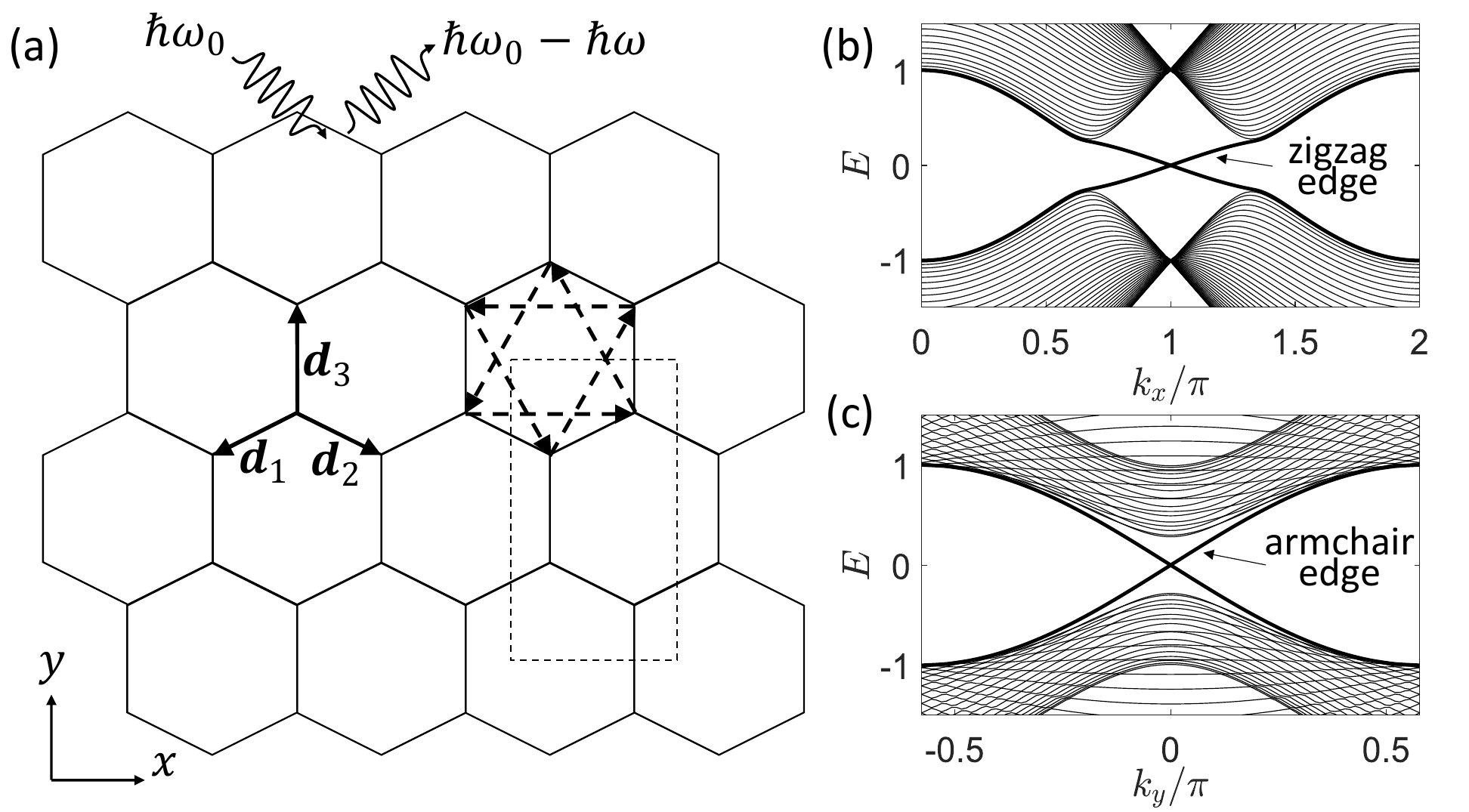}
	\caption{(a) The honeycomb lattice for the Kitaev model. The solid/dashed arrows denotes the nearest/next-nearest neighbours. The dashed rectangle denotes a unit cell used for numerical calculations. (b)/(c) The energy spectrum with open zigzag/armchair edges and $ J=2, K=0.1 $. The in-gap edge states are pointed out. }
	\label{fig:lattice}
\end{figure}

\paragraph{Quantum spin liquid ---} 
Consider the following Kitaev model Hamiltonian \cite{Kitaev}
\begin{align}
H = \sum_{i,\nu} J_\nu  \sigma_{\bm r_i}^{\nu} \sigma_{\bm r_i+\bm d_\nu}^{\nu} + h_\nu \sigma_{\bm r_i}^\nu ,
\label{h0} 
\end{align}
where $ \bm r_i $ is the displacement vector at site $ i $ on a honeycomb lattice as shown in FIG. \ref{fig:lattice}(a). Each site has three nearest neighbors connected by vectors $ \bm d_{\nu=1,2,3} $ respectively and $ \sigma^{\nu}_{\bm r_i} $ are the spin operators at site $ i $. 
When the external magnetic field $ h_\nu $ is small, the Hamiltonian for the low-energy sector can be written in the Majorana representation \cite{Kitaev,XT}
\begin{align}
H = \frac{iJ_\nu}{2}\sum_{\langle ij\rangle} c_i c_j + \frac{iK}{2} \sum_{\langle ij\rangle'} c_i c_j.
\label{h1}
\end{align}
Here  $ \langle ij\rangle  $  denotes nearest neighbours  connected by $ \bm d_{1,2,3} $ while  $ \langle ij\rangle' $  are next-nearest neighbours as indicated by the dashed arrows in FIG. \ref{fig:lattice}(a). 
$ K $ is determined by $ h_\nu $. Note that we have excluded the vortex excitations in Eq. (\ref{h1}) because they are gapped and we consider the energy regime inside the gap. Thus, our results are applicable  when the temperature is small compared to the vortex gap and the photon frequency $ \omega $ is smaller than this gap.

When $ h_\nu=0 $ and thus  $ K =0 $, there are two types of QSLs given by Eq. (\ref{h0}) --- a gapless phase when $ |J_1| \leqslant |J_2|+|J_3| ,|J_2| \leqslant |J_1|+|J_3|,|J_3| \leqslant |J_2|+|J_1|$, and gapped phases otherwise. Here we consider the case $ J_1=J_2=J_3=J $ and $ h_1 h_2 h_3 \neq 0$, then $ K \approx h_1 h_2 h_3/J  $ which makes the system gapped except on the edges where chiral Majorana modes exist. \cite{Kitaev} The energy spectra with open zigzag and armchair edge modes are shown in FIG.\ref{fig:lattice}(b) and (c) respectively.  

Spins can induce photon scattering through super-exchange coupling \cite{Fleury} and the Raman operator for the bond of two spins on the sites $ i $ and $ j $ is (up to an undetermined constant) \cite{Fleury,Shastry,Nasu,Knolle1}
\begin{align}
R_{ij}\sim  (\hat{\bm e}_{\text{in}}\cdot \bm d_{ij})  (\hat{\bm e}_{\text{out}}\cdot \bm d_{ij}) J^\nu_{ij} \sigma_i^\nu \sigma_j^\nu,
\label{R0}
\end{align}
where $ J^\nu_{ij} $ is the coupling between spins and $ \bm d_{ij} $ is the vector connecting them. The unit vector $ \hat{\bm e}_{\text{in/out}} $ is the polarization direction of the incident/outgoing light. 
In our case, it can be written in Majorana representation as
\begin{align}
 R_{ij} \sim   \frac{i J }{2} (\hat{\bm e}_{\text{in}}\cdot \bm d_{ij})  (\hat{\bm e}_{\text{out}}\cdot \bm d_{ij})   c_i c_j. 
 \label{R}
\end{align}
The Raman scattering intensity for a given photon frequency shift $ \omega $ is  
\begin{align}
I(q,\omega) = \int_{-\infty}^{\infty} dt e^{i\omega t} \langle   \hat{R}(q,t) \hat{R}(-q,0) \rangle.
\label{eq:I}
\end{align}
where $ \hat{R}(q,t) $ is the Fourier component along the $ x $-direction of Eq.(\ref{R}) in the Heisenberg picture.

We are interested in the quantity $ I(q,\omega) $ induced by the one-dimensional Majorana edge modes. Before a calculation in the full two-dimensional model, we study tentatively a one-dimensional 
effective edge theory with the Majorana Hamiltonian
\begin{align}
H_{\rm eff}
= \sum_{k>0} v k c_k^\dagger c_k.
\label{eq:heff}
\end{align}
And the one-dimensional version of Eq.(\ref{R}) is (in the reciprocal space) 
$R(q)  \sim v \sum_{k} \sin k c_{-k-q} c_k$  which becomes $R(q) =\sum_{k} v k c_{k+q}^\dagger c_k$
in the continuum limit. 
Substitution of this operator into Eq.(\ref{eq:I})  and application of the effective Hamiltonian $H_{\rm eff} $ lead to (using Wick's theorem and assuming the temperature $ T=0 $)
\begin{align}
	\langle   \hat{R}(q,t) \hat{R}(-q,0) \rangle \sim 
		\sum_{k,k'} k k'e^{-ivqt} [\delta_{k,-k'}+\delta_{k+q,k'} ], 
\end{align}
when $ 0<k<-q $ or $ 0<k'<q $,	and zero otherwise. $ \delta_{k,k'} $ is the Kronecker delta. Assuming infinite size and changing the wave vector summation to an integral, we obtain
$
	\langle   \hat{R}(q,t) \hat{R}(-q,0) \rangle \sim v^2 |q|^3 e^{i|vq|t}.
$
Thus, using Eq. (\ref{eq:I}),
\begin{align}
	I(q,\omega) \sim v^2 |q|^3 \delta (\omega+|v q|).
\end{align}
Assuming that the incident light has a Gaussian distributed intensity proportional to $ e^{-x^2/\Delta_x^2} $, the total Raman intensity is
\begin{align}
	I(\omega) = \int dq e^{-( q \Delta_x)^2/2} I(q,\omega) \sim  \frac{|\omega|^3 }{v^2} e^{-(\omega\Delta_x)^2/2v^2}
	\label{eq:I1d}
\end{align}
for $ \omega<0 $ and it vanishes when $ \omega>0 $.  The constant $ \Delta_x $ denotes the width of the light distribution. At small $ \omega $, we get $ I \sim |\omega|^3 $.  The factor $ 1/v^2 $ is related to the density of states, which is expected to affect the scattering strength. Eq. (\ref{eq:I1d}) has a maximum  of 
\begin{align}
I(\omega_0) \sim v/\Delta_x^3
\label{eq:Imax1}
\end{align}
at the frequency shift 
\begin{align}
\omega_0= \sqrt{3}v/\Delta_x.
\label{eq:Imax2}
\end{align} 
When $ \omega \ll \omega_0 $, $ I(\omega) $ is proportional to $ \omega^3$. 
$ I(\omega_0) $ decreases as $ \Delta_x $ increases because the Raman scattering process of the chiral Majorana modes requires translational symmetry breaking. 

\begin{figure}
	\includegraphics[width=3.0in]{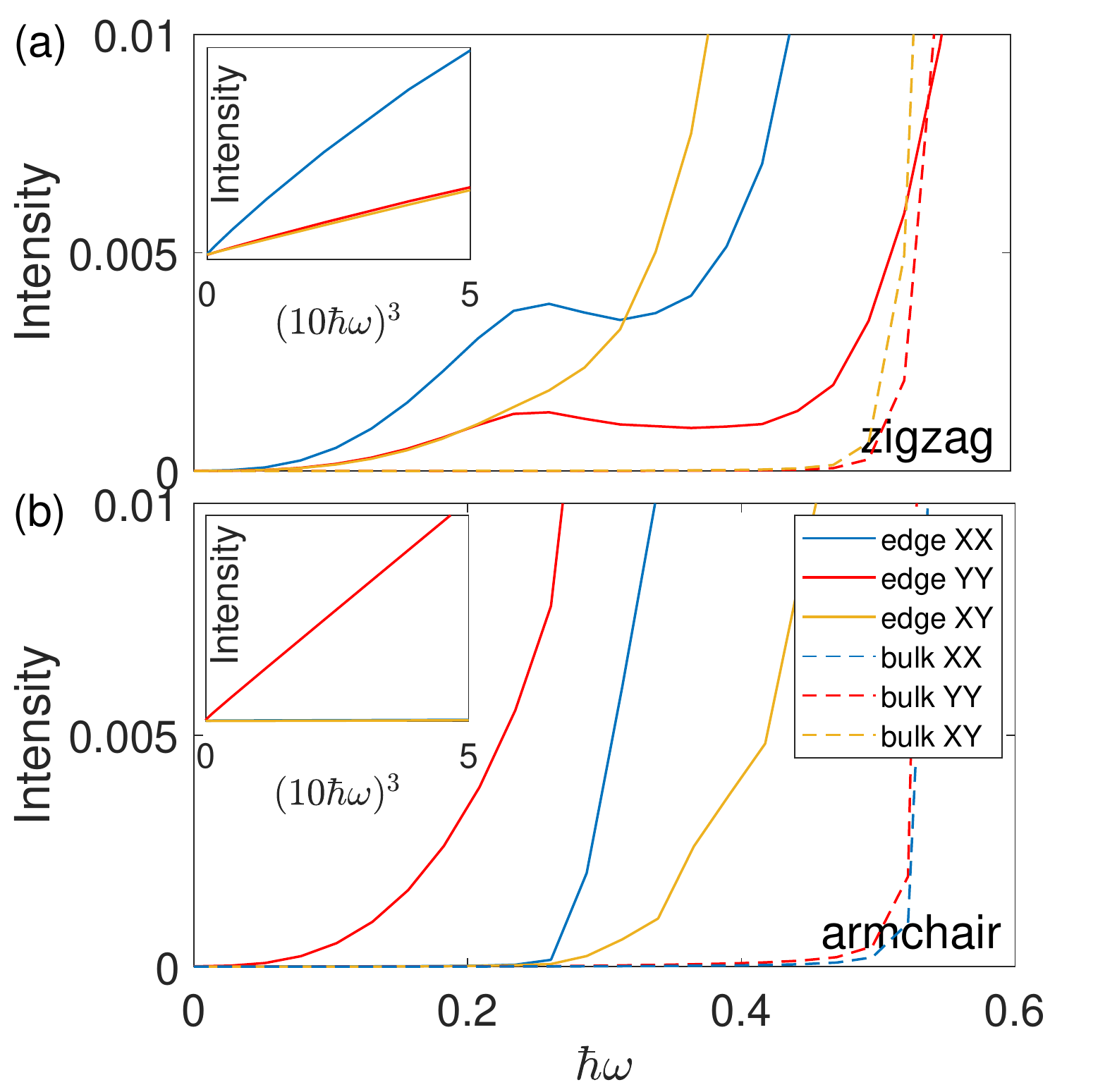}
	\caption{ The zero-temperature local Raman spectra for various choices of light-polarizations on (a) zigzag edges and (b) armchair edges. The solid/dashed curves are obtained with the detection spot  on the edges/in the bulk. The size of the detection spot is $ n_x\times n_y $, with $ n_x $ and $ n_y $ being the number of rectangular unit cells along $ x $- and $ y $-directions respectively. (a)$ n_x\times n_y=1\times 2 $. The number of sites along $ y $-direction is $ L_y=20 $. (b), $ n_x\times n_y =2\times 1 $. The number of sites along $ x $-direction is $ L_x=30 $. $ J=2, K=0.1$ in both (a) and (b). The symbols XX, YY and XY denote the polarizations of the incident (the first letter) and outgoing (the second letter) lights. }
	\label{fig2}
\end{figure}

Back to the two-dimensional Kitaev model, there are two typical kinds of edges when open boundaries are considered --- zigzag and armchair. Let us assume that the light shines on a small rectangular area of size $ n_x \times n_y $ ($ n_x $ and $ n_y $ denote the numbers of rectangular unit cells,  as shown in FIG. \ref{fig:lattice}(a), along $ x $- and $ y $-directions respectively.). Then the zero-temperature Raman intensity with various polarizations and at different positions (bulk or edge) are shown in FIG.\ref{fig2}. 

The solid (dashed) curves are obtained with the light-shining area on the edge (in the bulk). In the low-frequency regime, the Raman intensity is finite on the edges but vanishes in the bulk, manifesting the existence of the edge states.
A closer look at this regime shows that the intensity is proportional to $ \omega^3 $ (see the insets in FIG.\ref{fig2}), as indicated by previous one-dimensional effective model analysis. Thus, a local Raman spectroscopy measurement provides a valid method to detect the Majorana edge modes in the QSL model. 
The topological gap suggested by the bulk response is around $ 2E_g=0.5 $, consistent with the energy spectrum in FIG. \ref{fig:lattice}(b) and (c).  For $ \hbar\omega>E_g $, the bulk states is involved and the $ \omega^3 $ feature is gone. The Raman scattering in this regime is contributed by the optical transition between the edge and bulk states. In the case of zigzag edges, the band edge is at momenta (about $ \pi \pm 0.35\pi  $) away from the Majorana-edge-mode-crossing point $ \pi $, as seen in FIG. \ref{fig:lattice}(b). Since the Raman scattering with $ \hbar\omega  \gtrsim E_g $ comes from the transitions between edge modes with $ k_x \approx \pi $ and the bulk states on the band edge, a momentum mismatch exists, which suppresses the Raman intensity. This is the reason that the Raman intensity decreases just after $ \hbar \omega  \approx E_g $, resulting in a bump-like feature of the curve in FIG. \ref{fig2}(a). On the other hand, there is no momentum mismatch when the armchair edges are considered. Thus the Raman intensity in FIG. \ref{fig2}(b) is monotonically increasing. 
The symbols XX, YY and XY denote the polarizations of the incident (the first letter) and outgoing (the second letter) lights. The signals are the strongest when both the incident and outgoing lights are polarized along the edge, i.e. XX for zigzag edges and YY for armchair edges.

\begin{figure}
	\includegraphics[width=3in]{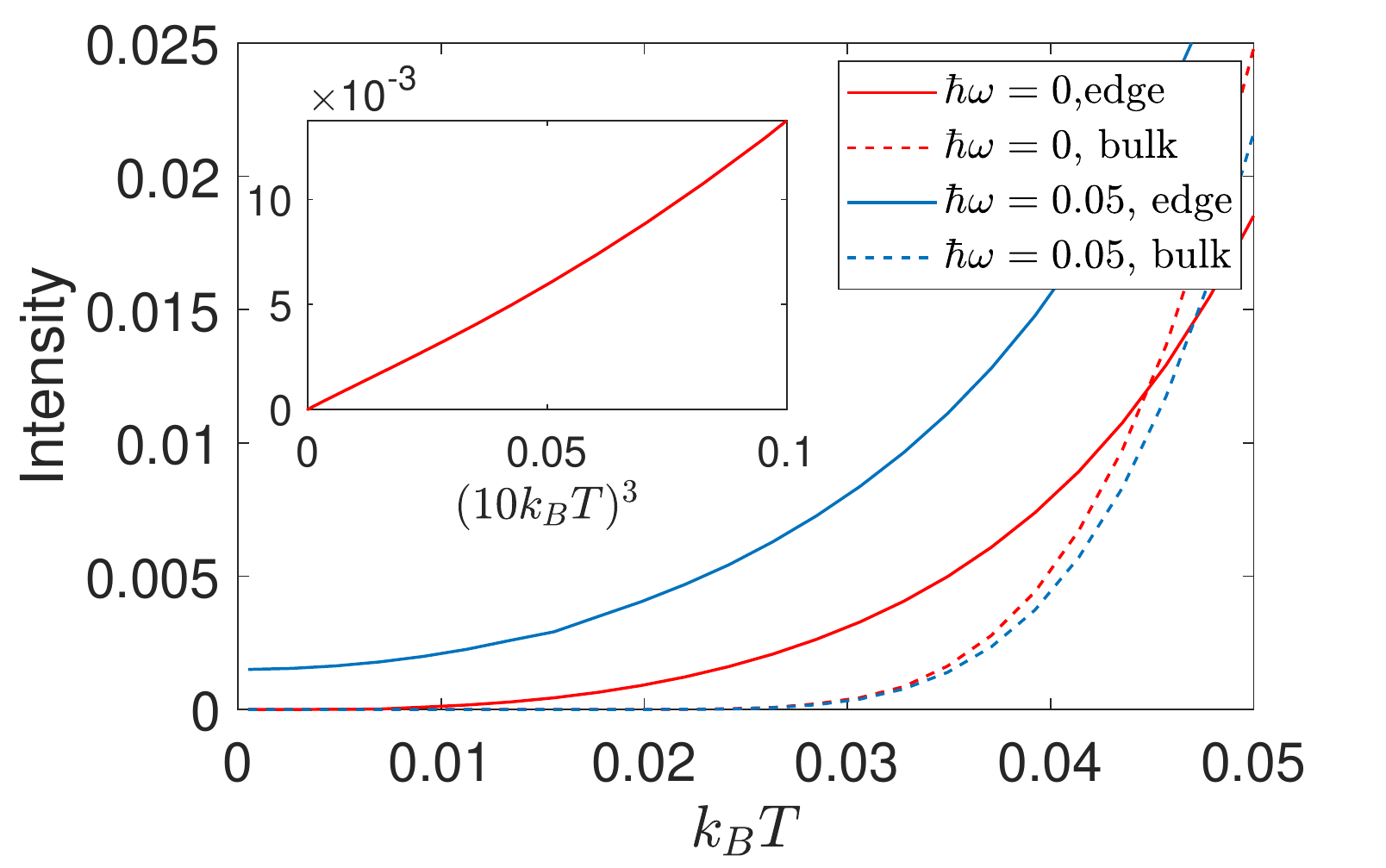}
	\caption{Temperature dependences of the local Raman intensities (XX polarizations) for various frequencies on the zigzag edge (solid curves) and in the bulk (dashed). The inset is the intensity on the edge at $ \omega=0 $ drew with the horizontal axis changed to $ (10k_B T)^3 $. }
	\label{fig_T}
\end{figure}

An exact calculation of finite temperature results needs to consider the complete Hilbert space including higher-energy sectors with vortex excitations. However, we limit ourselves to finite but small (compared to the vortex excitation gap) temperatures. Then the excitation of vortices is exponentially suppressed and the effect of temperature is reflected on the Fermi distribution functions of the low-energy Majorana states.   
In this way, we obtain the low-temperature behaviour as shown in FIG. \ref{fig_T}. For $ \omega=0 $, the intensity is proportional to $ T^3 $.

\paragraph{Topological superconductors ---}

One may expect similar Raman scattering features to appear with chiral Majorana edge modes in other systems such as topological superconductors (TSCs). To compare the Raman scattering by Majorana edge modes in QSLs and TSCs, let us consider a spinless $ p $-wave superconductor whose Hamiltonian is 
\begin{align}
	 H_{p}=\sum_{\bm k} {\xi(\bm k)} \psi_{\bm k}^\dagger \psi_{\bm k}  + \Delta [(\sin k_x + i \sin  k_y) \psi_{\bm k}^\dagger \psi_{-\bm k}^\dagger + h.c.] 
	 \label{eq:Hp}
\end{align} 
where the operator $ \psi_{\bm k} $ annihilates an electron of energy $ \xi(\bm k)=-2t[\cos(k_x)+\cos(k_y)]-\mu $.

\begin{figure}
	\includegraphics[width=3.0in]{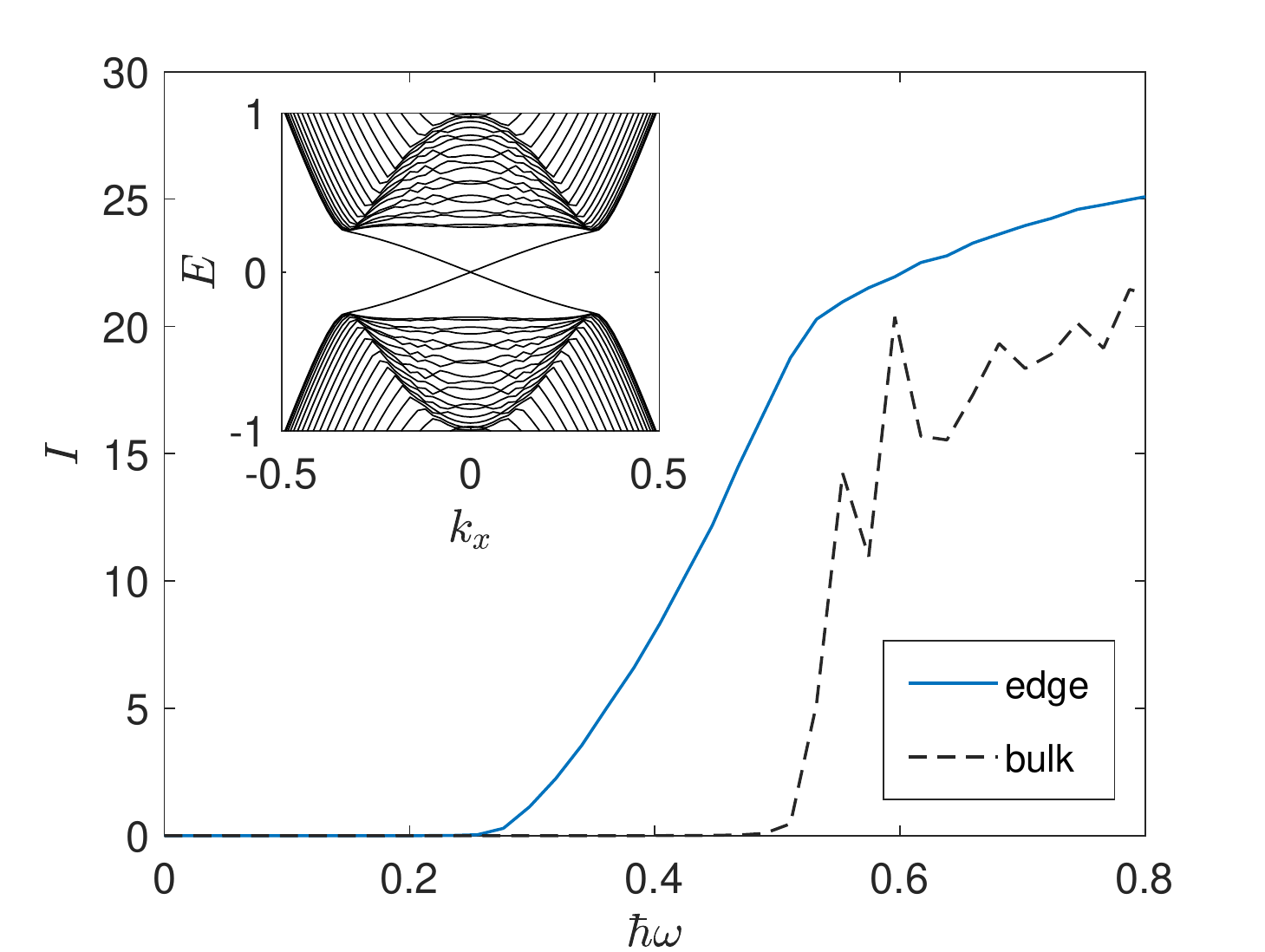}
	\caption{The Raman intensities in the bulk and on the edge of a spin-less $ p $-wave superconductor. The inset shows the energy spectrum. The size along the $ y $-direction is $ 80  $ sites. The parameters $ t=1, \mu=1, \Delta=0.3 $. }
	\label{fig3}
\end{figure}

The Raman operator in models of itinerant electrons is essentially given by the density operator.  In tight-binding models, it becomes \cite{Abrikosov}  (assuming $ \hat{\bm e}_\text{in} = \hat{\bm e}_\text{out}=\hat{x}  $ without loss of generality), 
\begin{align}
	R_p(\bm q) \sim  \sum_{\bm k} \psi^\dagger_{{\bm k}+\bm q} \frac{\partial^2 \xi(\bm k)}{\partial k_x^2} \psi_{\bm k},
	\label{eq:Re}
\end{align}
By rewriting Eqs. (\ref{eq:Hp}-\ref{eq:Re}) in the real-space representation along $ y $-direction and applying open boundary conditions there, we obtain the Raman intensity for this system using Eq. (\ref{eq:I}) as shown in FIG. \ref{fig3}. Both the spectrum in the inset and the bulk response indicate that the bulk energy gap is around  $ 2E_g = 0.54 $. The results for the two edges coincide   due to the obvious spatial inversion symmetry. Most importantly, no Raman scattering happens for frequencies $ \omega<E_g$. This is a remarkable difference from the results for the QSL model. 

The vanishing of the edge Raman scattering in TSCs may be understood by noting that the density operator, $ \rho(x) = \psi(x)^\dagger \psi(x) = \frac{1}{2}[1+i c(x)\tilde{c}(x)] $ (the normal fermion operator is expressed in terms of Majorana operators by $ \psi(x) = c(x)+i \tilde{c}(x) $), which also gives the Raman operator here, must involve two Majorana fields $ c $ and $ \tilde{c} $.  Since only one Majorana field exists below energy $ E_g $, Raman scattering cannot happen. When $ \omega >E_g  $, the bulk states get involved, with which another species of Majorana modes can be defined although they are not energy eigenstates. Thus, the Raman intensity becomes finite. The Raman response in the range $ E_g<\omega<2E_g $ is a result of the collaboration of edge and bulk states, similar to the QSL case. 

Comparing the Raman scattering in  the QSL and that in the TSC makes it clear that the electrodynamics of Majorana edge modes depends on their environments, i.e. on the bulk properties. This is related to the fact that Majorana modes do not preserve charges themselves. Charge conservation is recovered only when the whole system, including the bulk and the edges, is taken into account. Thus, is no general theories of the electrodynamics of Majorana edge modes. 

Although Majorana edge states alone in TSCs do not contribute to Raman scattering, they can absorb light and induce an optical conductivity (the real part) $ \Re[\sigma_{xx}] $ that is proportional to $ \omega^2 $ \cite{JJH2021}. In a Kitaev spin liquid, photon absorption involves the gapped vortex excitations and thus is negligible at the low-energy limit. \cite{Bolens} We obtain TABLE \ref{table1} as a comparison between QSLs and TSCs.

\begin{table}
	\begin{tabular}{|c|c|c|}
		\hline 
		& $\Re[\sigma_{xx}]$ & Raman  \\ 
		\hline 
		QSL	& 0  & $\sim \omega^3 $ \\ 
		\hline 
		TSC	& $ \sim \omega^2 $   &  0 \\ 
		\hline 
	\end{tabular} 
\caption{Raman scattering and real-part optical conductivity due to Majorana edge modes in QSLs and TSCs.}
\label{table1}
\end{table}

\paragraph{Conclusion and discussion--- }
We have shown that the chiral Majorana edge modes in the Kitaev model of QSL induce Raman scattering whose intensity is proportional to the frequency cube $ \omega^3 $ in the small-$ \omega  $ limit. For a finite temperature $ T$ that is small compared to the bulk gap, the Raman intensity is proportional to $ T^3 $ when $ \omega \rightarrow 0 $. 
The bulk system must be taken into account when electrodynamics of Majorana edge modes is considered. Majorana edge modes from different origins may lead to different electrodynamic properties. For instance, the Majorana edge modes alone in TSCs cannot induce Raman scattering while they can absorb light.  

A candidate material that shows the properties of the Kitaev model is $ \alpha $-RuCl$ _3 $ \cite{Sandilands,Jansa}. The spin-excitation gap of this material is about $\Delta_0=1  $meV \cite{Jansa}. The topological gap $ \Delta_1 $ opened by an in-plane magnetic field $ B $ is about $\Delta_1 \approx 4\times 10^{-3}  B^3 $meV$\cdot $ T$^{-3}$ \cite{Jansa}. If $ B=5 $T, then $ \Delta_1=0.5$meV and the Raman scattering purely due to the chiral Majorana edge modes should be detected with the photon energy shift $ \hbar\omega < 0.25 $meV (or the frequency shift $ \omega <0.38 $THz) and the temperature $ T \ll  10 $K.

Now we estimate the Raman intensity based on our numerical results.  Figure 1(b) indicates  $ v\sim 1/\Delta_1$. Thus, a stronger field and a smaller size results in larger Raman intensity according to Eq. (\ref{eq:Imax1}). 
Since $ J\approx 30 $meV \cite{Jansa}, the gap in our QSL model calculation as shown in FIG. \ref{fig2} is about $ 0.25J \approx 8 $meV. The real gap $ \Delta_1 =0.5meV$ is $ 16 $ times smaller and thus the real edge state velocity should be $ v_{real}=v_{model}/16 $. The peak position of FIG. \ref{fig2}(a) is about half of the gap, i.e. $ \hbar \omega_{0,model} \approx 4 $meV. 
The $ \Delta_{x} $ we used in our calculation is  $ \Delta_{x,model}\approx 1 $nm. 
According to Eq. (\ref{eq:Imax2}), a realistic peak position with $ \Delta_{x,real}=5nm $ is $ \hbar\omega_{0,real}\approx  \frac{1}{16\times5}\hbar \omega_{0,model}= 0.05$meV or $ \omega_{0,real} =  75$GHz. Since $ I(\omega_{0,model}) \approx 4\times 10^{-3} I_{bulk}$ according to FIG. \ref{fig2}(a), we obtain $ I(\omega_{0,real}) =\frac{1}{16\times 5^3}I(\omega_{0,model})  \approx 2\times 10^{-6} I_{bulk}$. 
As a reference, the bulk Raman intensity $ I_{bulk} $ in our model calculation is of the order unity, which corresponds to the value measured by Sandilands et al.\cite{Sandilands}. 
It may seem too small a signal to be observed if a factor of $ 10^{-6} $ is there. However, the gigantic local field generated by a scanning tunneling microscope (STM) tip can enhance the Raman scattering by a factor of $ 10^7$ or even larger \cite{Zhang2017,Verma2017} and thus the signal can be well detectable.

Real materials may be a stack of many layers instead of rather than exactly two-dimensional. In principle, the edge states of each layer may couple with each other and form  two-dimensional surface states, instead of the one-dimensional modes we have discussed here. However, a Heisenberg coupling between layers involves vortex excitations which are gapped by $ \Delta_0$. So the inter-layer coupling is irrelevant in the low-energy limit.

\begin{acknowledgments}
\paragraph{Acknowledgments --- }
N.N. was supported by Ministry of Education, Culture, Sports, Science, and Technology
Nos. JP24224009 and JP26103006, the Impulsing Paradigm
Change through Disruptive Technologies Program of Council
for Science, Technology and Innovation (Cabinet Office,
Government of Japan), and Core Research for Evolutionary
Science and Technology (CREST) No. JPMJCR16F1 and No. JPMJCR1874, Japan. JJH was supported by the RIKEN Incentive Research Projects. 	

\end{acknowledgments}

\bibliographystyle{apsrev4-1}

\end{document}